\documentclass[prd,letterpaper,showpacs,superscriptaddress,floatfix]{revtex4}

\usepackage{graphicx,psfrag,amsmath,amssymb,amsfonts,bbm,latexsym,color,dcolumn,bm}

\begin{document}

\title{On the use of the proximity force approximation for deriving
limits \\ to short-range gravitational-like interactions \\ from sphere-plane
Casimir force experiments}

\author{Diego A. R. Dalvit}

\affiliation{Theoretical Division, MS B213, Los Alamos National Laboratory, Los Alamos, NM 87545, USA}

\author{Roberto Onofrio}

\affiliation{Dipartimento di Fisica ``Galileo Galilei'',Universit\`a  di Padova, Via Marzolo 8, Padova 35131, Italy}

\affiliation{Department of Physics and Astronomy, Dartmouth College, 6127 Wilder Laboratory, Hanover, NH 03755, USA}
 
\date{\today}

\begin{abstract}
We discuss the role of the proximity force approximation in deriving
limits to the existence of Yukawian forces - predicted in the
submillimeter range by many unification models - from Casimir 
force experiments using the sphere-plane geometry. 
Two forms of this approximation are discussed, the first used in most 
analysis of the residuals from the Casimir force experiments
performed so far, the second recently discussed in this context in 
R. Decca {\it et al.}, [Phys. Rev. D \textbf{79}, 124021 (2009)]. 
We show that that the former form of the proximity force approximation 
overestimates the expected Yukawa force and  that the relative
deviation from the exact Yukawa force is of the same order of 
magnitude, in the realistic experimental settings, as the relative 
deviation expected between the exact Casimir force and the Casimir 
force evaluated in the proximity force approximation. This implies 
both a systematic shift making the actual limits to the Yukawa force weaker 
than claimed so far, and a degree of uncertainty in the
$\alpha-\lambda$ plane related to the handling of the various approximations 
used in the theory for both the Casimir and the Yukawa forces. 
We further argue that the recently discussed form for the proximity 
force approximation is equivalent, for a geometry made of a generic 
object interacting with an infinite planar slab, to the usual exact 
integration of any additive two-body interaction, without any need 
to invoke approximation schemes.
If the planar slab is of finite size, an additional source of 
systematic error arises due to the breaking of the planar translational 
invariance of the system, and we finally discuss to what extent this may
affect limits obtained on power-law and Yukawa forces.
\end{abstract}


\maketitle
\section{Introduction}

Several unification schemes merging gravity and the standard model 
of strong and electroweak interactions predict the existence of
short-range forces with coupling strength of the order of 
Newtonian gravity \cite{Giudice}. 
Efforts to evidence a {\sl fifth force} have been envisaged 
regardless of any concrete unification scheme since various decades
\cite{Fujii,Fischbach}, and there are compelling reasons to improve
our limits especially in the largely unexplored submillimeter range. 
Constraints in both coupling and range for these interactions have 
been obtained with various experimental setups, including the recent 
configurations using a disk-shaped torsional balance parallel to a 
rotating flat surface \cite{Adelberger1,Adelberger2,Adelberger3,Adelberger4}, 
or micromechanical resonators in a parallel plane geometry 
\cite{Price1,Carugno,Price2,Kapitulnik1,Kapitulnik2,Kapitulnik3}. 
Due to the surge of activity in the study of Casimir forces, limits
have been also given in the submicrometer range based on the level 
of accuracy between Casimir theory and experiment.  
However, unlike the case of experiments performed between bodies kept 
at larger distances, the use of the parallel plane geometry on such 
small lengthscale has been proven to be challenging in terms of 
parallelism \cite{Bressi1,Bressi2,Bressi3}, and therefore the
attention has been focused on the analysis of the residuals in the
Casimir theory-experiment comparison involving the sphere-plane configuration. 
 
Dedicated efforts to obtain limits from sphere-plane Casimir experiments
have involved the use of the so-called Proximity Force Approximation
(PFA) \cite{Fischbach1,Decca2003,Decca2005,Klim,Decca2007,Decca2007bis,MostepanenkoJPA},
which allows to map the force $F_{sp}$ between a sphere of radius $R$ and 
a plane located at a distance $a$ from the sphere 
into the energy per unit area $E_{pp}$ of the parallel plate 
configuration, namely $F_{\rm sp}(a)=2 \pi R E_{\rm pp}(a)$ \cite{Derjaguin}. 
This approximation is believed to be valid in the limit $a \ll R$ 
and to hold with a high degree of accuracy for forces 
between entities concentrated on the surfaces, such as electrostatic 
or Casimir forces between conductors \cite{Gies,Bordag,Krause}.  
Obviously, in order to test how well PFA approximates the exact 
force, one needs either to compute the interaction exactly or 
at least to assess reliable bounds. For the electrostatic 
sphere-plane interaction, the exact analytical result for the force 
is well-known and has a closed form \cite{Smythe}, such that deviations from PFA
can be readily analyzed. For the Casimir sphere-plane interaction, the 
exact force has been computed only very recently, both for ideal
\cite{Emig1,Emig2,Paulo2008}  and real metallic plates \cite{Canaguier}.
Available analytical and numerical 
results seem to indicate that, at least for zero temperature and within the 
used plasma model, deviations from PFA applied to the sphere-plane Casimir 
interaction are small, of the order of $0.1\%$ or higher, in recent Casimir 
experiments aiming to put limits to Yukawa interactions.

It has been argued in \cite{ReplyPRARC} that the application of the
PFA to forces acting between entities embedded in volumetric
manifolds, such as gravitational forces or their putative short-range 
components, is in general {\sl invalid} and has to be carefully 
scrutinized in each specific configuration. 
Based on this suggestion, a recent reanalysis of the PFA in the case
of gravitational and Yukawian forces has been discussed in
\cite{DeccaPFA}. The main conclusion of this reanalysis is that 
``a confusion with different formulations of the PFA'' existed in 
the previous literature, and that ``care is required in the
application of the PFA to gravitational forces''. 
This confusion is stated to originate from 
a specific form of the PFA used so far, to be contrasted 
with a more general formulation of the PFA. In \cite{DeccaPFA} it is 
also claimed that the difference between the two PFAs is negligible 
in the actual configuration used to give the allegedly best limits 
obtained in the 100 nm range \cite{Decca2007,Decca2007bis}. 

In this paper we further discuss the meaning of the PFA in the case 
of volumetric forces. We argue that the discrepancy between the two
forms of the PFA is a significant source of error in the determination 
of bounds on parameters of Yukawian forces from force residuals in Casimir
sphere-plane experiments performed so far  that used PFA to model
such non-Newtonian forces.  We then show that the general form for the
PFA discussed in \cite{DeccaPFA} is simply a different
choice of the infinitesimal volume for integrating the force due to an
extended object, and coincides with the exact result only in the case 
when one of the two surfaces is an infinite planar slab (or semispace). 
The level of approximation in using the two PFAs for Yukawa
forces in the sphere-plane geometry
is of the same  order of magnitude as the Casimir theory-experiment comparison, that
uses PFA to compute the sphere-plane Casimir force (as already noticed
in \cite{DeccaPFA}). Therefore, since such a
comparison provides force {\sl residuals} that are in turn compared
against the theory of Yukawa forces to obtain limits on its
$\alpha-\lambda$ parameter space, the use of these {\sl subsequent} PFA
approximations of comparable level of approximation provides 
a possible source of systematic error, not carefully accounted for so far.
We also argue that other volumetric effects not directly related
to the PFA, such as the finite size of the planar surface used in the 
actual experiments, may provide a source of systematic error not taken 
into account so far, which strongly affects the limits to power-law 
forces \cite{Buisseret2007, MostepanenkoJPA}, but should not be a major 
source of concern on limits to Yukawa forces. 
We believe that, considering the various number of complications
related to the sphere-plane geometry, upgraded versions of parallel plates experiments such 
as the ones discussed in \cite{Price1,Carugno,Price2,Kapitulnik1,Kapitulnik2,Kapitulnik3}
could provide limits on Yukawian and power-law forces in the
submicrometer range more immune to a set of systematic errors 
characteristic of the sphere-plane configuration.  


\section{Proximity Force Approximations and volumetric forces between
extended objects}
 
In order to introduce the notation and as a prelude to our discussion, 
we briefly summarize the results contained in \cite{DeccaPFA}.
The actual experimental configuration used in \cite{Decca2007} is not 
a parallel plate geometry, rather it is a sphere-plane geometry, and  
the PFA is used to map the force between a sphere and a plane
$F_\mathrm{sp}$ into the energy per unit area of the parallel plate
configuration $E_\mathrm{pp}$
\begin{equation}
F_\mathrm{sp}(a)=2 \pi  \bar{R} E_\mathrm{pp}(a) ,
\label{APPROXPFA}
\end{equation}
where $\bar{R}=\sqrt{R_x R_y}$ is the geometrical average of the 
principal radii of curvature of the spherical surface evaluated at 
its point of minimum distance from the plane. In the experiment
reported in \cite{Decca2007}, the force is measured by looking at the 
frequency shift of a mechanical resonator, as customary in atomic 
force microscopy \cite{Giessbl}, and as first reported in the
context of Casimir force measurements in \cite{Puppo}.
The frequency shift is proportional to the gradient of the force, 
and therefore
\begin{equation}
\Delta \nu^2= \frac{1}{4 \pi^2 m} \frac{\partial
F_\mathrm{sp}}{\partial a}=\frac{\bar{R}}{2 \pi m} 
\frac{\partial E_\mathrm{pp}}{\partial a}=
\frac{\bar{R}}{2 \pi m}P_\mathrm{pp},
\end{equation}
where $P_\mathrm{pp}$ is the plane-plane pressure, and $m$ is the mass of the resonator.
The measure of the frequency shift can then be mapped, via use of
Eq. (\ref{APPROXPFA}), into the equivalent pressure exerted between 
two fictitious parallel plates mimicking the actual sphere-plane geometry.  
Within the validity of Eq. (\ref{APPROXPFA}), this is a valid assumption 
for the case of forces acting between surfaces, such as electrostatic 
forces between conductors or Casimir forces. 

A first sign of the fact that there can be issues with the PFA in dealing with 
volumetric forces, such as the hypothetical Yukawian forces of 
gravitational origin, is manifested by noticing that the exact formula 
for the Yukawa force between two infinite parallel slabs depends on
the thicknesses of both slabs, which implies that the PFA formula applied to the volumetric
Yukawa force in the sphere-slab configuration also depends on {\sl both} thicknesses (and on the sphere
radius). However, the exact sphere-slab force obviously depends only
on the slab thickness and on the sphere radius - it does not, and
cannot, depend on the thickness of the {\sl metaphysical} slab
introduced in the virtual mapping to the parallel geometry. 
Indeed,  consider the Yukawa potential energy for two pointlike masses 
$m_1$ and $m_2$, located at positions ${\bf r}_1$ and ${\bf r}_2$ respectively,
\begin{equation}
U_{\rm Yu}({\bf r}_1,  {\bf r}_2)=- \alpha G m_1 m_2 
\; \frac{e^{-| {\bf r}_2- {\bf r}_1|/\lambda}}{| {\bf r}_2- {\bf r}_1|},
\end{equation}
where, as usual, the strength of the Yukawa interaction is
parameterized in terms of Newton's gravitational constant 
$G$ through a dimensionless quantity $\alpha$, and $\lambda$ is its range. 
Assuming that the Yukawa interaction is additive, once integrated over two infinite, homogeneous 
parallel slabs separated by a distance $a$, one derives the corresponding pressure $P_\mathrm{Yu}$,
\begin{equation}
P_\mathrm{Yu}(a) = - 2 \pi \alpha G \rho_1 \rho_2  \lambda^2
e^{-a/\lambda} (1-e^{-D_1/\lambda})(1-e^{-D_2/\lambda}) , 
\label{pressure}
\end{equation}
where $D_1$ and $D_2$ indicate the thickness of each slab, 
$\rho_1$ and $\rho_2$ their densities. 
The exact Yukawa interaction in the sphere-slab geometry can be
readily computed, assuming additivity. The result is \cite{DeccaPFA}
\begin{equation}
F^\mathrm{exact}_{\mathrm{Yu}}(a) =
-4 \pi^2 \alpha G \rho_1 \rho_2 \lambda^3 R e^{-a/\lambda}  
(1-e^{-D_1/\lambda})(1-\lambda/R+e^{-2R/\lambda}+
 e^{-2R/\lambda} \lambda/R ).
\label{YukExact}
\end{equation}
As we mentioned above, most recent experimental works on limits to extra-gravitational
forces from sphere-plane Casimir measurements used the usual PFA.
In this approximation, the Yukawa force between a homogeneous sphere and an infinite homogeneous
slab of thickness $D_1$ is
\begin{equation}
F^\mathrm{PFA}_{\mathrm{Yu}}(a)= 2 \pi R P_{\rm Yu}(a) = 
-4 \pi^2 \alpha G \rho_1 \rho_2 \lambda^3 R 
e^{-a/\lambda}(1-e^{-D_1/\lambda})(1-e^{-D_2/\lambda}).
\label{YukPFA}
\end{equation}
In this case one needed to consider, in order to map the actual sphere-plane
configuration into a parallel plate geometry, a fictitious upper 
plate of thickness $D_2$ large enough, {\it i.e.} much larger than the explored 
Yukawa range ($D_2 \gg \lambda$). Again, this situation may appear disturbing to whoever believes 
that any experiment-theory comparison should not rely on the
introduction of arbitrary parameters not having a tangible, 
measurable counterpart in the concrete experimental setup.
Clearly the PFA prediction Eq. (\ref{YukPFA}) fails to give the exact result 
Eq. (\ref{YukExact}) in the range of its supposed validity, $a \ll R$,
and it is necessary to assume $\lambda \ll R, D_2$ in order for PFA to
tend to the exact result. But in this limit the volumetric nature of 
the interaction is lost, since the atoms in the ``bulk" no longer 
contribute appreciably to the total force. Likewise, PFA fails to 
give the exact Newtonian interaction between the sphere and the slab,
even in the range of its supposed validity $a \ll R$.

\begin{figure}[t]
\begin{center}
\includegraphics[width=0.45\textwidth]{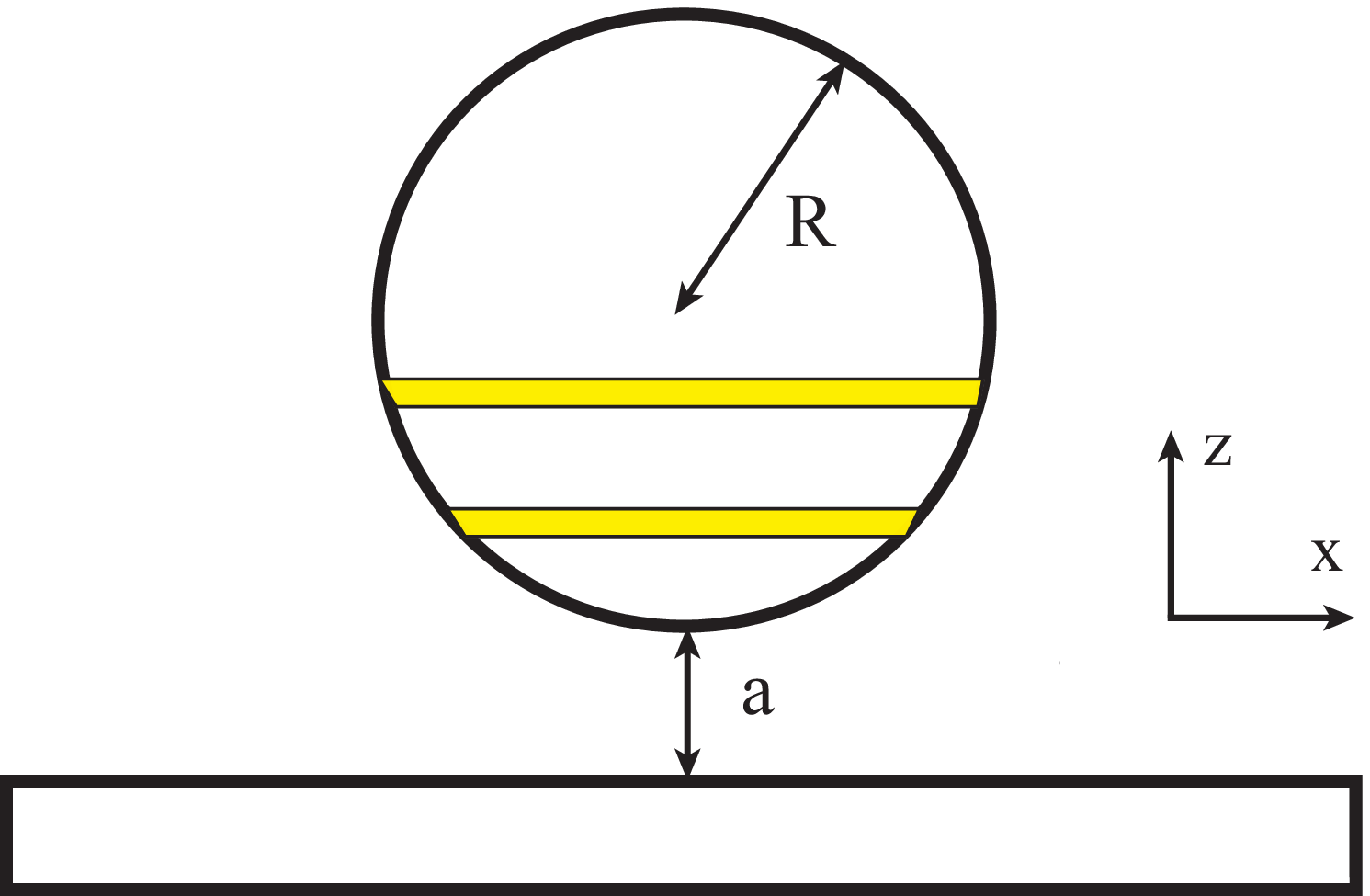}
\includegraphics[width=0.45\textwidth]{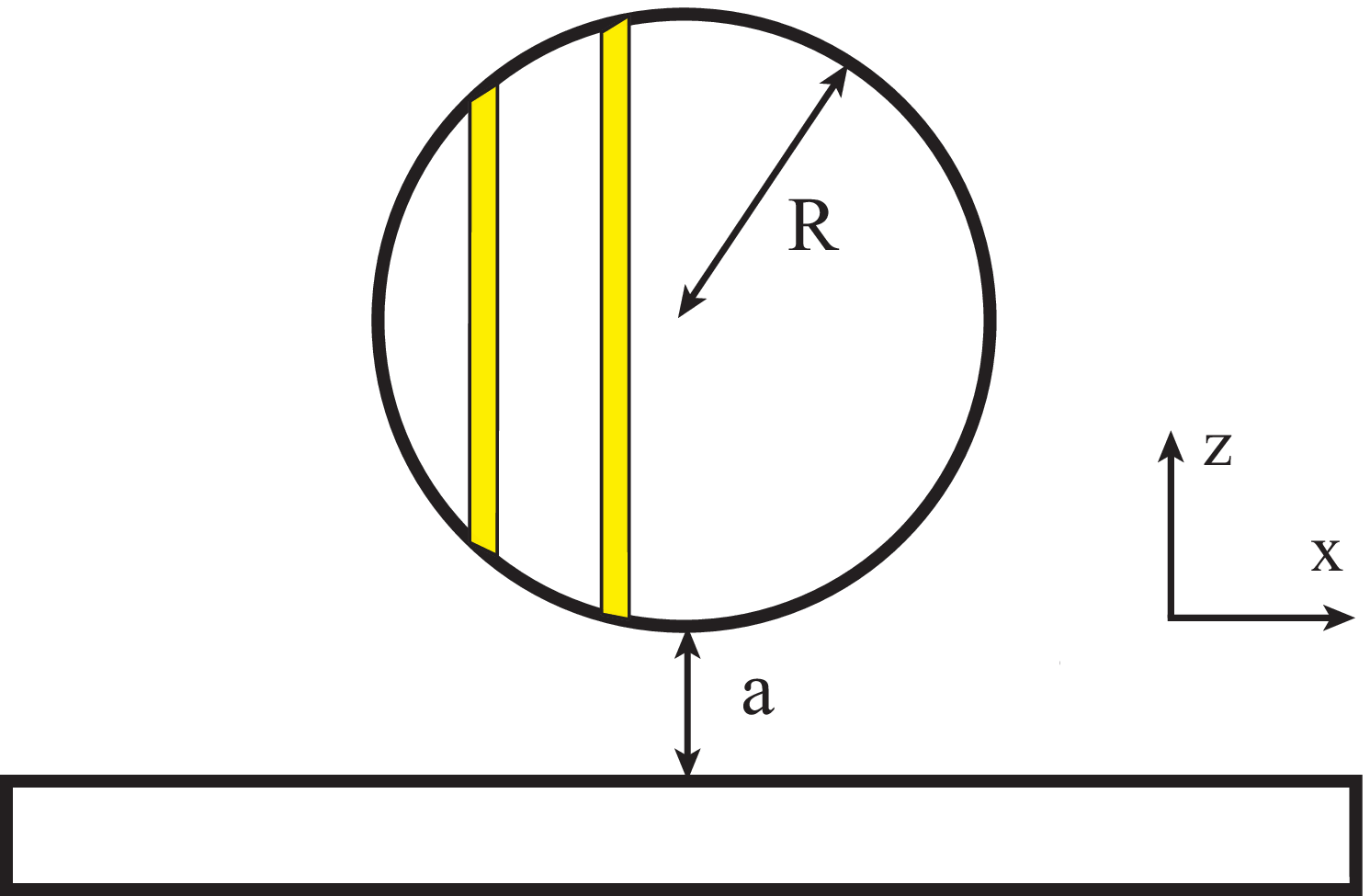}
\end{center}
\caption{Schematics of the integration for the sphere-slab
configuration according to the usual slicing along horizontal 
infinitesimal slabs used for the exact calculation (left),  
and slicing using vertical columns (right) as used in the EPFA calculation. 
As far as the surface facing the sphere is an infinite plane (and
therefore translational invariance of the potential $V$ due to the plane
is satisfied), these merely correspond to two different and equivalent 
choices for the integration volume.}  
\label{pfayuk.fig1}
\end{figure}

The authors of \cite{DeccaPFA} consider the most general formulation
of PFA \cite{Derjaguin1934}, that we will call ``exact" PFA formula (EPFA) to distinguish
it from the usual PFA approximation, described above.  In the EPFA
the force between two compact bodies is expressed as the sum of forces between plane
parallel surface elements $dx dy$. The $z$ component of the force is
\begin{equation}
F_z^\mathrm{EPFA}(a) = \int \int_\sigma dx dy  \; P(x,y,z(x,y)) ,
\end{equation}
where $P(x,y,z(x,y))$ is the pressure between two parallel plates 
at a local distance $z(x,y)=z_2(x,y)-z_1(x,y)>0$ ($z_i(x,y)$ 
are the surfaces of the two bodies),  $a$ is the distance between them (smallest value of
$z(x,y)$), and $\sigma$ is the part of the $(x,y)$-plane where both surfaces are defined (see Fig. 1).
The EPFA prediction for the Yukawa force between a sphere and an infinite planar slab is
\cite{DeccaPFA}
\begin{equation}
F^\mathrm{EPFA}_{\mathrm{Yu}}(a) =
-4 \pi^2 \alpha G \rho_1 \rho_2 \lambda^3 R e^{-a/\lambda}  
(1-e^{-D_1/\lambda})(1-\lambda/R+e^{-2R/\lambda}+
 e^{-2R/\lambda} \lambda/R ) , 
\label{YukEPFA}
\end{equation}
which coincides with the exact result of Eq.(\ref{YukExact}). In Fig.2 we plot the ratio $\eta$
\begin{equation}
\eta =\frac{F^{\rm EPFA}_{\rm Yu}}{F^{\rm PFA}_{\rm Yu}}=
\frac{1-\lambda/R+e^{-2R/\lambda}+e^{-2R/\lambda} \lambda/R}{1-e^{-D_2/\lambda}}
\label{equeta}
\end{equation}
as a function of the Yukawa parameter $\lambda$, for different values
of the sphere radius keeping fixed $D_2 \rightarrow \infty$ (left
plot), and for different values of $D_2$ keeping fixed the sphere 
radius at $R=150 \mu$m (right plot). 
Note that $\eta$ is independent of the sphere-slab
separation $a$. When $D_2 \gg  \lambda$,
as surely realized in the left plot,  PFA always {\sl overestimates}
the EPFA result, i.e. $\eta<1$ (similarly to how the PFA overestimates 
the exact Casimir force in the sphere-plane geometry). Note also that when
$D_2 \gg \lambda$  the atoms in the ``bulk" of the two bodies do not 
contribute appreciably to the Yukawa force, thereby making it
effectively of a surface character (i.e., non-volumetric), as in the 
case of Casimir or electrostatic forces.
Instead, for values of $D_2 \simeq 10 \lambda$ (or smaller) the
volumetric character of the Yukawa interaction is manifest,
and $\eta$ is no longer less than one (right plot). 
In this case the PFA applied to Yukawa forces is invalid and 
in the data analysis one should at least declare the value 
of $D_2$ imagined for which the limits are assessed.
Overestimating the Yukawa force leads to stronger limits for 
the coupling constant $\alpha$ for a given $\lambda$ with respect to 
the proper use of the EPFA. Considering the relatively small margins 
of improvement  reported recently (see for instance Fig. 3 in
\cite{Decca2007}) a systematic shift due to the use of the PFA 
instead of the EPFA may lead to significant changes for the exclusion 
region in the $\alpha-\lambda$ plane.   
Both plots show that the use of PFA instead of EPFA is unreliable 
especially in the region near or above $\lambda$=100 nm. 
Unfortunately the region in between 100 nm and few $\mu$m 
is also the one {\sl directly} explored with the Casimir 
force experiments, since actual measurements take place in this 
range of distance between the involved objects. It is known that the 
best limits on Yukawa interactions can be set for $\lambda$ of the 
order of the actual explored distance between the two bodies $\simeq
a$, and the extrapolation of the measurements to smaller $\lambda$ 
is affected by the fast growth of the bounds as $\alpha \propto \exp(a/\lambda)$.  

The fact that the EPFA Eq. (\ref{YukEPFA}) gives the correct exact results for 
the Yukawa (and also gravitational) force in the sphere-plane
configuration is in fact a trivial consequence of the additivity
of these interactions  and of the translational invariance of the
infinite plane surface, the shape of second surface (in this case a
sphere) being irrelevant.  Indeed, the EPFA is just a {\sl different}
parameterization of the exact formula of addition of forces between 
particles. On one hand, the exact interaction energy between a test 
mass and an infinite slab (or half-space) depends only on the normal
coordinate $z$, being independent of the in-plane coordinates $x, y$
by symmetry. Hence, the potential due to the infinite slab is 
$V(x,y,z)=V(z)$. For a body ({\it e.g.} a sphere) of mass density 
$\rho(x,y,z)$, additivity implies that the total interaction energy 
can be obtained as
\begin{equation}
U_{\mathrm{body}}= \int dx dy dz \; \rho(x,y,z) \; V(z),
\label{body}
\end{equation}
and similarly for the force. Since $V$ depends only on $z$, it is 
convenient to compute the integral by adding forces at different 
slices at constant $z$, {\it i.e.} considering infinitesimal 
slices in $z$, then evaluating first the potential energy of a 
slice of the body parallel to the plane at a distance $z$ 
\begin{equation} 
W(z)=\int dx dy \; \rho(x,y,z) \; V(z), 
\end{equation}  
and then integrating along $z$ the quantity $W(z)$ one obtains the exact
expression for the body-plane interaction $U_{\mathrm{body}}$ (see Fig. 1 left). 
On the other hand, the EPFA states that the interaction energy between 
the body and the plane is obtained from slicing the body into
cylinders perpendicular to the plane (see Fig. 1 right) and integrating  the
cylinder-plane interaction energy along the portion of the body that
faces the plane (i.e., one must integrate over the surface $\sigma$ on
the plane that is the normal shadow of the body). The potential 
energy of this column of the body centered around $(x,y)$ is
\begin{equation}
G(x,y)=\int dz \; \rho(x,y,z) \; V(z),
\end{equation}
and then integrating along $x,y$ the quantity $G(x,y)$  one gets the EPFA expression
for the body-plane interaction  $U_{\mathrm{body}}$ (see Fig. 1 right).
Again, since the interaction is additive and it does not depend on the $x,y$
coordinates, this integral is exactly equal to the previous one: 
we are simply integrating the same function $U(z)$ over the body 
using a different parameterization of the volumetric integral. 
The same holds for the force between any body (not necessarily a sphere) and an infinite slab
(see also \cite{Emig}). 

\begin{figure}[t]
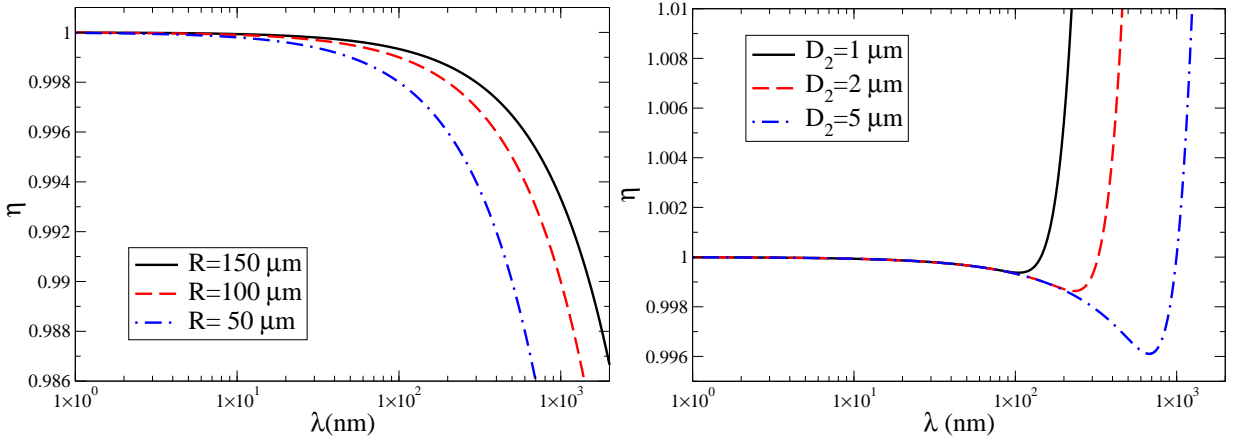

\begin{center}
\includegraphics[width=0.45\textwidth]{pfayuk.fig2a.eps}
\includegraphics[width=0.45\textwidth]{pfayuk.fig2b.eps}
\end{center}
\caption{(Color online) Comparison between the EPFA and the PFA 
versus the range of the Yukawian force for a homogeneous sphere above an infinite 
homogeneous slab.
(Left) Plot of $\eta$ versus $\lambda$ 
in the case of spheres of radius $R=$150, 100, and 50$\mu$m, 
in the limit of $D_2 \to \infty$.
(Right) Same plot but for a sphere of radius $150 \mu$m and 
finite values of the metaphysical parameter $D_2$ representing  
the thickness of the upper slab introduced in \cite{DeccaPFA} 
for the comparison of the actual sphere-plane geometry to 
the parallel plate case using the PFA.}
\label{pfayuk.fig2}
\end{figure}

However, when none of the two bodies is an infinite slab (e.g., two
spheres of radii $R_i$, mass densities $\rho_i(x,y,z)$, separated by a
distance $a$ along the $z$ direction),  translational invariance along
$x-y$ is obviously broken, and the EPFA does not coincide with 
the exact formula.  The exact interaction energy $U(x,y,z)$ between 
a source body and a test mass at position $(x,y,z)$ can be easily
computed. For instance, for the spherical source body $U$ depends 
only on $r=(x^2+y^2+z^2)^{1/2}$, with the origin of coordinates
at the center of the sphere. Integrating $U(x,y,z)$ over the volume
of the second body one gets the exact result. For example, for the
two-spheres case the gravitational energy depends only on the
center-to-center distance and scales as $1/a$. 
Let us compare this known exact result with the EPFA prediction. 
One slices each body in cylindrical slabs, calculates the slab-slab 
interaction energy $U_{ss}(z)$ that depends on the local distance 
$z$ between the slabs, and finally one adds up these contributions 
over the shadow of one of the bodies on the other one. 
It is clear that EPFA cannot give the exact result since $U_{ss}$ 
is translational invariant but the exact $U$ is not, and EPFA fails 
to predict the exact $1/a$ dependency. 
Therefore the EPFA formula for the 
energy and force of additive two-body interactions is a trivial 
reparameterization of the exact result when one of the bodies is 
an infinite plane (or slab). For other geometries, EPFA fails 
to give the correct result as a consequence of the broken 
translational invariance. In particular, this is the case 
of a sphere above a finite size slab, like in experiments, 
especially those involving slabs of typical sizes comparable 
to those of the sphere.

In the experimental configuration used in \cite{Decca2007} various
substrates are present on the sphere and on the slab.  
Imagining that the layered slab is infinitely long, the Yukawa 
potential at a distance $z$ from the top layer due to the slab is
 \begin{eqnarray}
 V^{\Delta}_{\rm Yu}(z) &= & - 2 \pi \alpha G \lambda^2 e^{-z/\lambda}  \left[
 \rho''_1 e^{- \Delta''_1/\lambda}  (e^{\Delta''_1/\lambda}-1) +
 \rho'_1 e^{- (\Delta''_1+\Delta'_1)/\lambda}  (e^{\Delta'_1/\lambda}-1) + \right. \nonumber \\
 && ~~~~~~~~~~~~~~~~~~~~~~~~
 \left. \rho_1 e^{- (\Delta''_1+\Delta'_1+D_1)/\lambda}  (e^{D_1/\lambda}-1) \right] .
 \label{testmass}
 \end{eqnarray}
Here $\Delta''_1$ and $\rho''_1$ are the thickness and density of the 
top layer, $\Delta'_1$ and $\rho'_1$ are the thickness and density of 
the middle layer, and $D_1$ and $\rho_1$ and the thickness and density 
of the lower part of the layered slab. The last factor in
Eq.(\ref{testmass}) can be considered a sort of
{\sl effective} density of the planar surface, in which the various 
densities are weighted by their thicknesses in units of $\lambda$ 
(indeed yielding their arithmetic average in the case of 
$\Delta', \Delta'', D_1 \ll  \lambda$).  

We can compute the exact expression for the Yukawa interaction energy
between the layered infinite slab and a layered sphere of mass 
density $\rho_2(x,y,z)$ using Eq. (\ref{body}). As discussed above,
this exact computation will trivially coincide with the EPFA expression. 
Let $R$ and $\rho_2$ be the radius and density of the sphere,
$\Delta'_2$ and $\rho'_2$ the width and density of the inner 
layer on the sphere, $\Delta''_2$ and $\rho''_2$ the width and 
density of the outer layer, and $a$ the distance from the outer 
layer of the sphere to the top of the layered slab. The total EPFA Yukawa
interaction energy can be written as a sum of contributions 
from each layer on the sphere, 
$U^{\Delta, \rm EPFA}_{\rm Yu}= U^{\Delta}_2 + U'^{\Delta}_2 +
U''^{\Delta}_2$, where
\begin{eqnarray}
U^{\Delta}_2 &=&  2 \pi \rho_2 \int_0^R r^2 dr \int_0^{\pi} d\theta \sin\theta
\rho_2 V^{\Delta}_{\rm Yu}(z),  \nonumber \\
U'^{\Delta}_2 &=& 2 \pi \rho_2' \int_R^{R+\Delta'_2}  r^2 dr \int_0^{\pi}
d\theta \sin\theta \rho'_2 V^{\Delta}_{\rm Yu}(z), \nonumber \\
U''^{\Delta}_2 &=& 2 \pi \rho_2'' \int_{R+\Delta'_2}^{R+\Delta'_2+\Delta''_2} 
r^2 dr \int_0^{\pi} d\theta \sin\theta \rho''_2 V^{\Delta}_{\rm Yu}(z). \nonumber
\end{eqnarray}
Note that, instead of using horizontal or vertical slicings for the volume integration as done in 
Fig. 1 for the non-layered case, we use spherical slicings more appropriate for the layered
sphere case. Here $z=a+\Delta''_2+\Delta'_2+R - r \cos\theta$ denotes the vertical 
position of any infinitesimal mass element inside the layered sphere.
Computing these integrals we obtain the EPFA expression for the Yukawa 
interaction energy between the layered infinite slab and the layered sphere
\begin{eqnarray}
&& U_{\rm Yu}^{\Delta, {\rm EPFA}}(a) = - 4 \pi^2 \alpha G \lambda^4 R \; 
e^{-(a+\Delta''_2+\Delta'_2)/\lambda} \nonumber \\
&& \times
\left\{
\rho''_1 e^{- \Delta''_1/\lambda}  (e^{\Delta''_1/\lambda}-1) +
\rho'_1 e^{- (\Delta''_1+\Delta'_1)/\lambda}  (e^{\Delta'_1/\lambda}-1) +
\rho_1 e^{- (\Delta''_1+\Delta'_1+D_1)/\lambda}  (e^{D_1/\lambda}-1)
\right\} \nonumber \\
&& \times
\left\{
\rho_2
\left[
1-\frac{\lambda}{R} + e^{-2 R/\lambda} + \frac{\lambda}{R}  e^{-2 R/\lambda}
\right] \right. + \nonumber \\
&& ~~~~~
\rho'_2
\left[
\left( 1-\frac{\lambda}{R} \right)  (e^{\Delta'_2/\lambda}-1) + \frac{\Delta'_2}{R} e^{\Delta'_2/\lambda}  +
e^{-2 R/\lambda}
\left(
\left(1- \frac{\lambda}{R} \right)  (1-e^{-\Delta'_2/\lambda}) + \frac{\Delta'_2}{R} e^{-\Delta'_2/\lambda}
\right)
\right] + \nonumber \\
&& ~~~~~
\rho''_2
\left[
\left( 1-\frac{\lambda-\Delta'_2}{R} \right) e^{\Delta'_2/\lambda}
(e^{\Delta''_2/\lambda}-1) + 
\frac{\Delta''_2}{R} e^{(\Delta'_2+\Delta''_2)/\lambda} + 
\right. \nonumber  \\
&&  ~~~~~~~~~
\left.
\left.
e^{-2 R/\lambda}
\left(
\left( 1+\frac{\lambda+\Delta'_2}{R} \right) e^{-\Delta'_2/\lambda}
(e^{-\Delta''_2/\lambda}-1) +  
\frac{\Delta''_2}{R} e^{-(\Delta'_2+\Delta''_2)/\lambda}
\right)
\right]
\right\} .
\label{UdeltaEPFA}
\end{eqnarray}
The EPFA expression for the corresponding force is 
$F^{\Delta,\rm EPFA}_{\rm Yu} = - \partial  U_{\rm Yu}^{\Delta,\rm EPFA} / \partial a=
 \lambda^{-1} U_{\rm Yu}^{\Delta, \rm EPFA}$. 
Note that when there are no layers on the slab ($\Delta'_1=\Delta''_1=0$) 
and no layers on the sphere ($\Delta'_2=\Delta''_2=0$), then 
the expression for the force that follows from Eq. (\ref{UdeltaEPFA})
is identical to Eq. (\ref{YukExact}).
\begin{figure}[t]
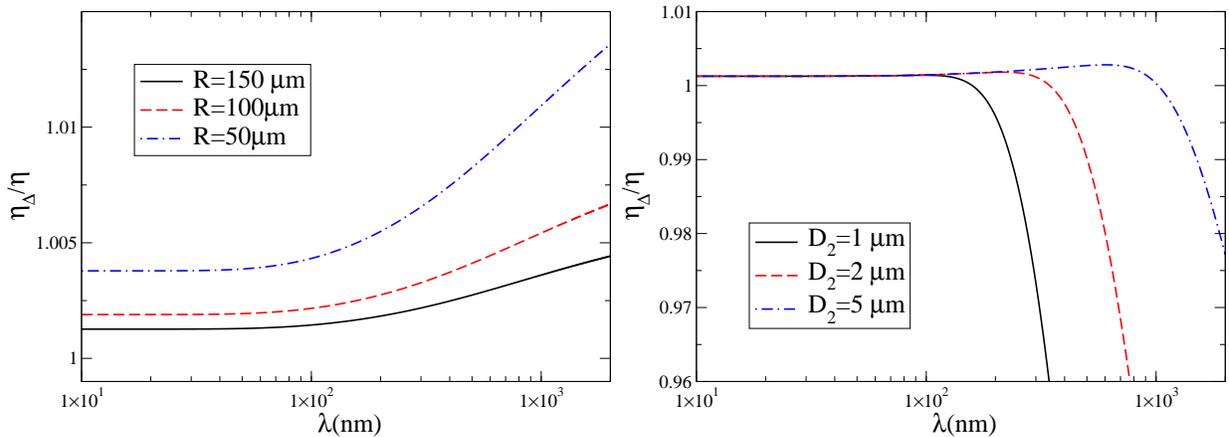

\begin{center}
\includegraphics[width=0.45\textwidth]{pfayuk.fig3a.eps}
\includegraphics[width=0.45\textwidth]{pfayuk.fig3b.eps}
\end{center}
\caption{(Color online)  
Ratio $\eta_{\Delta}/\eta$  for the
comparison of the EPFA and PFA schemes for the multilayered and
corresponding homogeneous situation (obtainable by using Eqs. (14) and
(15) with $\Delta_2'=\Delta_2''=0$, $\rho_2'=\rho_2''=0$, and 
replacing $R$ with $R+\Delta_2'+\Delta_2''$).      
(Left) Ratio $\eta_{\Delta}/\eta$ versus the range of the Yukawian force for different values of
the radius of the inner sphere. The parameters for the layered sphere are
$\Delta_2'=10$ nm, $\Delta_2''=180$ nm, $\rho_2$=4.1 g/cm${}^3$, 
$\rho_2'$=7.14 g/cm${}^3$, and $\rho_2''$=19.28 g/cm${}^3$.
The parameters for the layered slab are
$D_1=3.5 \mu$m,  $\Delta_1'=10$ nm, $\Delta_1''=210$ nm, $\rho_1$=2.33 g/cm${}^3$,
$\rho_1'$=7.14 g/cm${}^3$, and $\rho_1''$=19.28 g/cm${}^3$.
In the evaluation of the PFA force $F^{\Delta, \rm PFA}_{\rm Yu}$, 
a value of the metaphysical parameter 
$D_2 = 10^8\mu$m is used. (Right) Ratio $\eta_{\Delta}/\eta$ versus the 
range of the Yukawian force for different values of the metaphysical
parameter $D_2$ and a radius of curvature of $R=150 \mu$m.}
\label{pfayuk.fig3}
\end{figure}
On the other hand, the PFA expression for the force between the 
layered infinite slab and the layered sphere is 
$F^{\Delta, \rm PFA}_{\rm Yu} = 2 \pi R P^{\Delta}_{\rm Yu}$, where 
$P^{\Delta}_{\rm Yu}$ is the pressure between two parallel 
layered slab, one identical to the previous slab, and a 
metaphysical  slab of width $D_2$ and density $\rho_2$, covered 
by two layers of widths and densities identical to the ones of the 
layered sphere above. Using Eq. (\ref{pressure}) for the various 
pairs of layers in the different slabs, we calculate the PFA 
expression for the layered sphere-slab force
\begin{eqnarray}
 &&
 F^{\Delta, \rm PFA}_{\rm Yu}(a) = - 4 \pi^2  \alpha G \lambda^3 R e^{-a/\lambda} \times \nonumber \\
 &&
 \left\{
 \rho_1 (1-e^{-D_1/\lambda})
 \left[
 \rho_2 e^{-(\Delta'_1+\Delta''_1+\Delta'_2+\Delta''_2)/\lambda} (1-e^{-D_2/\lambda}) +
 \rho'_2 e^{-(\Delta'_1+\Delta''_1+\Delta''_2)/\lambda} (1-e^{-\Delta'_2/\lambda}) + 
 \right. \right. \nonumber \\
&& ~~~~~~~~~~~~~~~~~~~~~~~
\left. \left.
\rho''_2 e^{-(\Delta'_1+\Delta''_1)/\lambda} (1-e^{-\Delta''_2/\lambda})
\right] + \right. \nonumber \\
 && ~~
 \rho'_1 (1-e^{-\Delta'_1/\lambda})
 \left[
 \rho_2 e^{-(\Delta''_1+\Delta'_2+\Delta''_2)/\lambda} (1-e^{-D_2/\lambda}) +
 \rho'_2 e^{-(\Delta''_1+\Delta''_2)/\lambda} (1-e^{-\Delta'_2/\lambda}) +
 \rho''_2 e^{-\Delta''_1/\lambda} (1-e^{-\Delta''_2/\lambda})
 \right] + \nonumber \\
 && ~
 \left.
 \rho''_1 (1-e^{-\Delta''_1/\lambda})
 \left[
 \rho_2 e^{-(\Delta'_2+\Delta''_2)/\lambda} (1-e^{-D_2/\lambda}) +
 \rho'_2 e^{-\Delta''_2/\lambda} (1-e^{-\Delta'_2/\lambda}) +
 \rho''_2 (1-e^{-\Delta''_2/\lambda})
 \right] 
 \right\} .
\label{UdeltaPFA}
 \end{eqnarray}
To assess the effect of the multilayered structures, we have evaluated the ratio
$\eta_{\Delta}/\eta$ with $\eta_{\Delta}= F^{\Delta, \rm EPFA}_{\rm Yu} / F^{\Delta, \rm PFA}_{\rm Yu}$  
and $\eta =  F^{\rm EPFA}_{\rm Yu} / F^{\rm PFA}_{\rm Yu}$, as a
function of $\lambda$ for three radii of curvatures of the sphere, assuming for the PFA calculation
a value of the metaphysical parameter $D_2= 10^8\mu$m (see Fig. 3 left). The effect of multilayers
is to slightly flatten $\eta_{\Delta}$ as compared to $\eta$ in the
homogeneous case (Fig. 2 left). The dependence of the same ratio for a
fixed value of $R$ and different values of the metaphysical parameter $D_2$ is shown in Fig. 3 right.
Note that $\eta_{\Delta}$ is independent of the sphere-slab separation, just as $\eta$ is.

As discussed in the Introduction, the PFA used in all recent
sphere-plane Casimir experiments for the Casimir theory-experiment
comparison is expected to approximate the exact Casimir force within
0.1 $\%$. This expectation comes from recent analytical approaches to the sphere-plane 
Casimir interaction \cite{Emig1,Emig2,Paulo2008,Canaguier} that, although formally exact,
require  the evaluation of the determinant of an infinite-dimensional matrix,
which becomes a numerically demanding task, especially in the PFA 
regime, $a \ll R$, where larger and larger matrices are needed for convergence. 
Numerical computations of the exact, zero-temperature sphere-plane Casimir force using 
parameters for metallic spheres ($R=10 \mu$m and optical response modeled by the simple 
plasma model with plasma wavelength $\lambda_p=136 $nm) show that deviations from
PFA can be as large as $20\%$ for the smallest $a/R \approx 0.5$ studied numerically 
(see Fig.2 of \cite{Canaguier}). An extrapolation to smaller values of $a/R$  
using a cubic polynomial fit of the numerical data is also provided in \cite{Canaguier}. 
Assuming one can use it for the recent Casimir sphere-plane experiment 
\cite{Decca2007} (with a radius of curvature $R=151.3 \mu$m), gives 
a deviation from PFA of the order of $0.1\%$ at the smallest value 
of $a/R \approx 0.001$ reached in the experiment ($a_{\rm min}  \approx 160$nm). 
Since the limits to non-Newtonian forces are obtained using 
the {\sl residuals} in the Casimir theory-experiment comparison, in order 
to meaningfully replace the exact formula of the Yukawa force with 
its PFA approximation, the level of accuracy between these two 
should be therefore a small fraction, for instance 10 $\%$, of 
the accuracy with which the Casimir force is controlled by
using PFA rather than the exact expression for the sphere-plane Casimir force.
If this condition is not fulfilled, the derived limits could 
be off also by a large, order of 100$\%$, correction. 
However, targeting a $10 \%$ accuracy level with respect to the 
Casimir theory-experiment accuracy implies deviations from $\eta=1$ 
of  $0.01 \%$, which can be obtained, as seen in Fig. 2, only in the 
range of $\lambda$ below 100 nm. The presence of substrates with 
different densities tends to mitigate the discrepancy between the 
EPFA and the PFA, as seen by the curves in Fig. 3, but there is an 
irreducible systematic factor even at small $\lambda$. Indeed, in 
the limit $\lambda \rightarrow 0$, we have
$\eta_{\Delta} \approx 1+(\Delta_2'+\Delta_2'')/R$, 
that, in the case of the experiment reported in \cite{Decca2007}, is 
equal to 1.00126, {\it i.e.} a correction already equal to 0.126 $\%$. 

All these systematic sources of uncertainty could be even larger in 
experiments for which the radius of curvature of the sphere 
is not adequately optimized. Indeed, the use of spheres with 
smaller radius of curvature is affected more by this effect, 
as emphasized in the left plot of Fig. 2 and in Fig. 3 for 
the cases of $R=50 \mu$m and $100 \mu$m. 
Moreover, for small spheres the PFA approximation to the Casimir 
force itself is less accurate. The use of spheres with large radius of
curvature is beneficial to reduce these sources of error in the 
experiment-theory comparison, but may face experimental issues 
recently identified in \cite{Kim1} and interpreted as due to
deviations from an ideal spherical geometry (as proposed in
\cite{DeccaComment}) and/or a consequence of larger sensitivity to 
electrostatic patch effects \cite{Kim1,ReplyPRARC}.


\section{Breaking the x-y translational invariance}
  
In the previous section we have seen that translational invariance is
crucial to make the EPFA reproduce the exact result, but in actual
experiments such an invariance is obviously satisfied only approximately,
leading to an additional source of systematic error related 
to the finite size of the surfaces, as we discuss here for 
both power-law forces, and for Yukawa forces. 
Instead of computing the more involved problem of the gravitational force
between a sphere and a finite-size slab, we consider here the simpler case of 
the gravitational force acting on a point-like test mass $m_2$ above the center of 
a disk of thickness $D_1$, radius $R_d$, and mass density $\rho_1$. We obtain
\begin{eqnarray}
F_g (z_2) & =& -2 \pi G \rho_1 m_2 \int_{-D_1}^0 dz_1 \int_0^{R_d} r dr
\frac{z_2-z_1}{[r^2+(z_2-z_1)^2]^{3/2}} 
\nonumber \\  
& = & -2\pi G \rho_1 m_2 \{ D_1 +{(R_d^2+z_2^2)}^{1/2}-{[R_d^2+(z_2+D_1)^2]}^{1/2} \} ,
\end{eqnarray}
where $z_2$ is the distance between the test mass and the disk.
The force becomes independent of $R_d$ only in the limit $R_d \gg D_1, z_2$
(in which case it is also independent of $z_2$).
In order to assess the different forces acting on the various parts of
a sphere in the presence of a disk of finite radius we evaluate the
ratio between the forces exerted at the point of
the sphere closest to the plane ($z_2=a$) and the farthest point  
($z_2=a+2R$). This quantity is simple to evaluate yet provides 
a practical figure of merit for how much the extended geometry of 
the sphere is affected by the finite size of the disk.
This gives a ratio $\xi_g=F_g(z_2=a)/F_g(z_2=a+2R)$:
\begin{equation}
\xi_g= \frac{\beta+[\gamma^2+\kappa^2]^{1/2}-[(\gamma+\beta)^2+\kappa^2]^{1/2}}
{\beta+[(2+\gamma)^2+\kappa^2]^{1/2}-[(2+\gamma+\beta)^2+\kappa^2]^{1/2}},
\end{equation}
where we have defined  $\beta \equiv D_1/R$, $\gamma \equiv a/R$,
and $\kappa \equiv R_d/R$.

\begin{figure}[t]
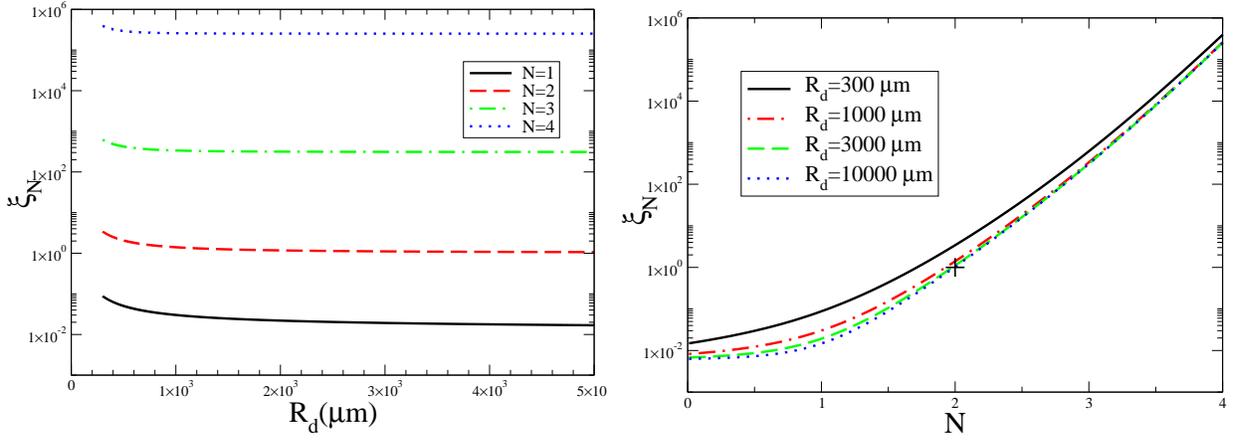

\begin{center}
\includegraphics[width=0.45\textwidth]{pfayuk.fig4a.eps}
\includegraphics[width=0.45\textwidth]{pfayuk.fig4b.eps}
\end{center}
\caption{(Color online) (Left) Plot of the ratio $\xi_N$ between the force 
$F(r) \propto r^{-N}$ exerted at the closest point of the sphere 
from the disk and the force exerted at the farthest point, 
versus the radius of the disk $R_d$ constituting the planar surface 
of finite size, for four different exponents $N$. We assume 
a radius of the sphere equal to $R=150 \mu$m, a sphere-plane
distance of $a$=100 nm, and that the center of the sphere is right above the center
of the disk.  Only the case of $N=2$ (Newtonian gravitation) 
gives a ratio of unity in the large $R_d$ limit.
(Right) Plot of the ratio $\xi_N$ versus the exponent of the 
power force law $N$ for various radii of the disk $R_d$.
The cross indicates the ratio $\xi=1$ for the case 
of an exponent $N=2$ and an infinite plane;  it is provided as a help to the eye 
to better show the convergence in the case of $N=2$ of $\xi_N$ to unity
with disks of progressively larger radii.}
\label{pfayuk.fig4}
\end{figure}

This is a large correction, of the order of 300$\%$, if a disk of 
radius equal to twice the radius of the sphere ($R_d=2 R$), in a 
geometrical setting not dissimilar from the one used in \cite{Decca2007}, 
is considered. Since the experiment in \cite{Decca2007} is anyway 
insensitive to the gravitational force, like any experiment
performed in the micrometer range, this is not a major practical
concern. However, in the case of a more generic power law 
such as $F_N=-K \rho_1 m_2/r^N$ we get:
\begin{equation}
F_N(z)= \frac{2 \pi K \rho_1 m_2}{(N-1)(N-3)} 
\{(z+D_1)^{3-N}-z^{3-N}+(R_d^2+z^2)^{(3-N)/2}-[R_d^2+(z+D_1)^2]^{(3-N)/2} \} ,
\end{equation}
apart from the cases of $N=1$ and $N=3$ in which logarithmic
integrations occur. In these two cases one obtains
\begin{equation}
F_{N=1}(z)=  \frac{\pi K \rho_1 m_2}{2} 
\{(z^2+R^2) \ln(z^2+R_d^2)-[(z+D_1)^2+R_d^2] \ln[(z+D_1)^2+R_d^2]+(z+D_1)^2 \ln (z+D_1)^2-z^2 \ln z^2 \} ,
\end{equation}
which in the limit $R_d \rightarrow \infty$ becomes independent of $z$, and
\begin{equation}
F_{N=3}(z)= - \frac{\pi K \rho_1 m_2}{2} \ln \left[\frac{(z^2+R_d^2)(z+D_1)^2}{z^2 [(z+D_1)^2+R_d^2]}\right] ,
\end{equation}
which in the limit $R_d \rightarrow \infty$ behaves as $\ln(1 + D_1^2/z^2)$.
Notice that the fact that the force is independent on the distance 
from an infinite plane is only characteristic of forces scaling 
with the inverse square of the distance, such as the gravitational force, 
making the integration of the force trivially geometrical. 
In the general case $n \neq 2$, even in the situation of a sphere 
in front of an infinite plane, different points of the sphere will
feel different forces, with the farthest point feeling smaller
(larger) force for a power law exponent larger (smaller) than 2, 
as a consequence of the interplay between the solid angle and 
the distance scaling of the force, which makes peculiar the 
$N=2$ case as expressed by the Gauss law. 
This is shown in Fig. 4 (left) for the cases of $N=1,2,3,$ and $4$, with 
the ratio $\xi_N$ between the forces evaluated at the top 
and  at the bottom of the sphere, and in Fig. 4 (right) by showing 
the same ratio versus the power law exponent for different values 
of the radius of the disk. Therefore, when considering power-law 
forces as the ones discussed for instance around Eq. 2 in 
\cite{MostepanenkoJPA}, one should then take carefully into account 
the finite size of the plane in deriving limits to these forces 
\cite{Note}.

\begin{figure}[t]
\begin{center}
\includegraphics[width=0.5\textwidth]{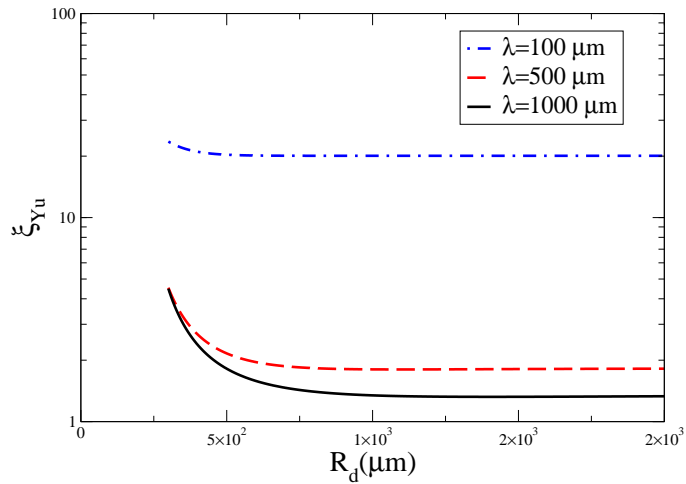}
\end{center}
\caption{(Color online) Plot of the ratio $\xi_{\rm Yu}$ between the 
Yukawa force exerted at the closest point of the sphere from the disk 
and the force exerted at the farthest point, versus the radius of the 
disk $R_d$ constituting the planar surface of finite size, for three different 
values of the Yukawa range $\lambda$. As before, we assume a radius of 
the sphere equal to $R=150 \mu$m, a sphere-plane distance 
of $a$=100 nm, and the center of the sphere right above the center
of the disk. The values of $\lambda$ are chosen to be 100 $\mu$m, 
500 $\mu$m, and 1000 $\mu$m. In the last case the force may be
considered as a long range one and the farthest point on the sphere 
is also contributing almost as the closest one. For $\lambda$ 
progressively smaller than the radius of the sphere the ratio $\xi_{\rm Yu}$ 
gets larger and larger and the dependence on $R_d$ is not appreciable.}
\label{pfayuk.fig5}
\end{figure}

Finally, we discuss the effect of the finite size of the planar
surface in the case of Yukawa forces. 
The potential energy of a pointlike particle of mass $m_2$ located 
at height $z$ along the axis of a planar disk surface of radius 
$R_d$, density $\rho_1$, and thickness $D_1$, is
\begin{eqnarray}
U_{\rm Yu}(z) & = & - \alpha G \rho_1 m_2 \int_0^R dr ~ r \int_0^{2\pi} 
\int_{-D_1}^0 [(z_2-z_1)^2+r^2]^{1/2}
e^{-\sqrt{(z_2-z_1)^2+r^2}/\lambda}
\nonumber \\
&=& -2 \pi \alpha G \rho_1 m_2 \lambda^2
\left[e^{-z/\lambda}(1-e^{-D_1/\lambda})-e^{-\sqrt{z^2+R_d^2}/\lambda}+e^{-\sqrt{(z+D_1)^2+R_d^2}/\lambda}\right] ,
\end{eqnarray}
and the related force is
\begin{eqnarray}
F_{\rm Yu}(z) & = & -\frac{\partial U_{\rm Yu}}{\partial z}= 
-2 \pi \alpha G \rho_1 m_2 \lambda e^{-z/\lambda}
(1-e^{-D_1/\lambda})  \nonumber \\
& & 
\times
\left\{1 - \frac{z}{R_d} \frac{e^{-[1+(z/R_d)^2]^{1/2}R_d/\lambda+z/\lambda}}{[1+(z/R_d)^2]^{1/2}(1-e^{-D_1/\lambda})}
+ \frac{z+D_1}{R_d} \frac{e^{-[1+((z+D_1)/R_d)^2]^{1/2}R_d/\lambda+z/\lambda}}{[1+((z+D_1)/R_d)^2]^{1/2}(1-e^{-D_1/\lambda})}
\right\} ,
\end{eqnarray}
\noindent
where the finite size terms appear as corrections to the indefinite
plane formula originating by the first term alone. 
As before, we introduce as figure of merit the ratio 
$\xi_{\rm Yu}=F_{\rm Yu}(z_2=a)/F_{\rm Yu}(z_2=a+2R)$. This ratio is 
very large for realistic configurations, expressing the short-range 
nature of the force. Indeed, even in the infinite plane limit we 
have a ratio of $\xi_{\rm Yu}=e^{2R/\lambda} \simeq e^{3000}$ in the case 
of a sphere of radius $R=150 \mu$m at a $\lambda=0.1 \mu$m. 
The dependence on the disk radius become significant only at values of 
$\lambda$ comparable to the radius of the sphere, as shown in Fig. 5.
The presence of suppression factors for the farthest point of 
the form $e^{2R/\lambda}$ makes very insensitive the Yukawa force 
to the finite size of the disk (for previous considerations, see 
also \cite{Bordag1998}). 


\section{Conclusions}

Our analysis, although confirming some of the results already
discussed in \cite{DeccaPFA}, draws quite different conclusions from 
the common outcome. In particular, we argue that the application of 
PFA to volumetric forces is not rigorous if considered in
its original formulation applied so far to compute the sphere-plane 
Yukawa interactions in the most sphere-plane Casimir force experiments 
reported in \cite{Decca2003,Decca2005,Klim,Decca2007,Decca2007bis,MostepanenkoJPA}. 
It application to volumetric forces is instead of trivial nature if
considered in the exact  formulation EFPA discussed in
\cite{DeccaPFA}, since the latter is identical to the exact
calculation, just differing in the choice of the 
infinitesimal integration volume. 
We have shown that the usual PFA is an invalid approximation to compute 
volumetric forces. In particular, it does not reproduce in its usual 
range of validity exact known expressions for gravitational and Yukawa 
interactions in non-translational invariant geometries, such as sphere-finite 
size slab or sphere-sphere configurations. The ``exact" PFA is the exact 
expression for any additive two-body interaction when one of the bodies 
is translational invariant, as is the case for a sphere in front of an 
infinite homogeneous slab or half-space. For non-translational invariant 
geometries, EPFA also fails to give the exact result for volumetric 
interactions, even in the regime of parameters where it is assumed to be valid,
therefore also being an invalid approximation for volumetric forces.

The difference between the two formulations of the PFA is shown 
to affect significantly the limits obtained so far unless one 
considers a regime of Yukawa range so small, $\lambda \ll R, D_2$,  that the 
approximation of a surface force ({\it i.e.} neglecting
the Yukawa force due to the atoms in the ``bulk'' of the 
two bodies, therefore manifestly of non-volumetric character) 
holds. By using the PFA the Yukawa force is overestimated and therefore 
the limits in the $\alpha-\lambda$ plane become more stringent than 
by using the exact force estimated via EPFA. Moreover, the use of 
PFA instead of the EPFA for the parameters of the experiment 
supposed to provide the strongest limits to Yukawian interactions 
\cite{Decca2007} occurs with an accuracy of the same order of 
magnitude with which the exact Casimir force is expected 
to be also approximated by the corresponding PFA. 
On one hand, since the Casimir theory-experiment comparison
provides force {\sl residuals} that are in turn compared
against the theory of Yukawa forces to obtain limits on them, 
the use of these {\sl subsequent} PFA
approximations of comparable level of approximation provides 
a possible source of systematic error, not carefully accounted for so far.
On the other hand, both Casimir and Yukawa PFAs 
overestimate the respective exact forces, and therefore 
the systematic source of error introduced by their use might be less critical than 
expected on first principles. In any case, it is important to 
assess as much as possible both sources of errors or, in 
alternative, to use exact expressions for the Yukawa and Casimir forces.
Furthermore, a systematic source of error present even in the EPFA
scheme for finite-size planar surfaces has been discussed and 
shown to be significant only for power-law forces. 
Our analysis suggests that future limits (or reanalysis 
of experiments already performed) on Yukawian forces 
should rely upon the use of the exact expression for 
the Yukawa force, as performed in \cite{Bordag1998,Masuda}.

It is also worth to point out that all the above considerations hold 
provided that the simple scenario of additive forces is assumed,
which is valid in general for weak forces among atoms in the low density 
limit, such that correlations leading to fluctuating forces are negligible. 
However, the hypothetical Yukawa forces should be located in 
a regime of coupling constants intermediate between the gravitational 
(additive) force and the Casimir (non additive) force. It is not 
understood a priori if the Yukawa force is weak enough to make 
the additivity assumption reliable, and this should be kept in 
mind in future broad-range searches of these forces.

Finally, considering the complications emerging in the sphere-plane geometry 
due to the presence of previously unidentified systematics such as 
the sensitivity to deviations from the ideal spherical geometry 
\cite{Kim1,Iannuzzi,DeccaComment,ReplyPRARC} and possible effects 
of variability of the contact potential with distance 
\cite{Speake,Stipe,Kim1,Iannuzzi,Kim2,Kimpatch}, it may be worth to focus future 
Casimir experiments to set bounds on extra-gravitational 
forces on the actual parallel-plane geometry, without the 
drawbacks of a virtual mapping from the sphere-plane
geometry made explicit in this paper. The stronger force signal 
expected for the same distance between the two surfaces, the 
reduced sensitivity to distance-dependent contact potentials due 
to image charges, the absence of deviations from a uniform radius 
of curvature, the existence  of exact mode summation techniques to 
compute the Casimir force, the possibility to control parallelism using 
recently developed technology \cite{PPtechnology}, and the 
possibility to compensate off-line the lack of parallelism 
by using the PFA as discussed in \cite{Bordagpp}, all point 
in the direction to continue this class of experiments in 
the actual parallel plane configuration, extending below the 10 $\mu$m 
range the results of the experiments described in 
\cite{Adelberger1,Adelberger2,Adelberger3,Adelberger4,Price1,Carugno,
Price2,Kapitulnik1,Kapitulnik2,Kapitulnik3}.



\begin{thebibliography}{99}

\bibitem{Giudice} 
S. Dimopoulos and G. F. Giudice, Phys. Lett. B \textbf{379}, 105 (1996).

\bibitem{Fujii} 
Y. Fujii, Nature \textbf{234}, 5 (1971); 
Ann. Phys. \textbf{69}, 494 (1972); Phys. Rev. D \textbf{9} 874
(1974); Int. J. Mod. Phys. A \textbf{6}, 3505 (1991).  

\bibitem{Fischbach} E. Fischbach and C. L. Talmadge, 
{\it The Search for Non-Newtonian Gravity} (AIP/Springer-Verlag, New York, 1999).

\bibitem{Adelberger1} 
E. G. Adelberger , B. R. Heckel, and A. E. Nelson, 
Ann. Rev. Nucl. Part. Sci. \textbf{53}, 77 (2003).

\bibitem{Adelberger2} 
D. J. Kapner, T. S. Cook, E. G. Adelberger,
J. H. Gundlach, B. R. Heckel, C. D. Doyle, and H. E. Swanson, 
Phys. Rev. Lett. \textbf{98}, 021101 (2007).

\bibitem{Adelberger3} 
E. G. Adelberger, B. R. Heckel, S. Hoedl, C. D. Doyle, D. J. Kapner, and A. Upadhye, 
Phys. Rev. Lett. \textbf{98}, 131104 (2007).

\bibitem{Adelberger4} 
S. Schlamminger, K.-Y. Choi, T. A. Wagner, J. H. Gundlach, and E. G. Adelberger, 
Phys. Rev. Lett. \textbf{100},  041101 (2008).

\bibitem{Price1} J. C. Price, in {\sl Proceedings of the International
Symposium on Experimental Gravitational Physics}, edited by
P.F. Michelson (World Scientific, Singapore, 1988), p. 436.

\bibitem{Carugno} G. Carugno, Z. Fontana, R. Onofrio, and C. Rizzo, 
Phys. Rev. D \textbf{55}, 6591 (1997).

\bibitem{Price2} J. C. Long, H. W. Chan, A. B. Churnside, E. A. Gulbis, 
M. C. M. Varney, and J. C. Price, Nature \textbf{421}, 922 (2003).

\bibitem{Kapitulnik1} 
J. Chiaverini, S. J. Mullin, A. A. Geraci, D. M. Weld, and A. Kapitulnik, 
Phys. Rev. Lett \textbf{90}, 151101 (2003).

\bibitem{Kapitulnik2} 
S. J. Smullin, A. A. Geraci, D. M. Weld, J. Chiaverini, S. Holmes, and A. Kapitulnik, 
Phys. Rev. D \textbf{72}, 122001 (2005).

\bibitem{Kapitulnik3} 
A. A. Geraci, S. J. Smullin, D. M. Weld, J. Chiaverini, and A. Kapitulnik, 
Phys. Rev. D \textbf{78}, 022002 (2008).

\bibitem{Bressi1} G. Bressi, G. Carugno, R. Onofrio, and G. Ruoso, 
Class. Quantum Grav. \textbf{17}, 2365 (2000).

\bibitem{Bressi2} G. Bressi, G. Carugno, A. Galvani, R. Onofrio, G. Ruoso, and F. Veronese, 
Class. Quantum Grav. \textbf{18}, 3943 (2001).

\bibitem{Bressi3} G. Bressi, G. Carugno, R. Onofrio, and G. Ruoso, 
Phys. Rev. Lett. \textbf{88}, 041804 (2002).

\bibitem{Fischbach1} E. Fischbach, D. E. Krause, V. M. Mostepanenko, and M. Novello, 
Phys. Rev. D \textbf{68}, 116003 (2003).

\bibitem{Decca2003} 
R. S. Decca, D. L\'opez, E. Fischbach, G. L. Klimchitskaya, D. E. Krause, and V. M. Mostepanenko, 
Phys. Rev. Lett. \textbf{91}, 050402 (2003).

\bibitem{Decca2005} 
R. S. Decca, D. L\'opez, E. Fischbach, G. L. Klimchitskaya, D. E. Krause, and V. M. Mostepanenko, 
Ann. Phys. (N.Y.) \textbf{318}, 37 (2005).

\bibitem{Klim} 
G. L. Klimchitskaya, R. S. Decca, E. Fischbach, D. E. Krause, D. L\'opez, and V. M. Mostepanenko, 
Int. J. Mod. Phys. A \textbf{20}, 2205 (2005).

\bibitem{Decca2007} 
R. S. Decca, D. L\'opez, E. Fischbach, G. L. Klimchitskaya, D. E. Krause, and V. M. Mostepanenko,
Phys. Rev. D \textbf{75}, 077101 (2007)

\bibitem{Decca2007bis}
R. S. Decca, D. L\'opez, E. Fischbach, G. L. Klimchitskaya, D. E. Krause, and V. M. Mostepanenko,
Eur. Phys. J. C \textbf{51}, 963 (2007).

\bibitem{MostepanenkoJPA} 
V. M. Mostepanenko, R. S. Decca, E. Fischbach, 
G. L. Klimchitskaya, D. E. Krause, and D. L\'opez, J. Phys. A \textbf{41}, 
164054 (2008).

\bibitem{Derjaguin} 
B. V. Derjaguin and I.I. Abrikosova, Sov. Phys. JETP {\bf 3},
819 (1957); B.V. Derjaguin, Sci. Am. \textbf{203}, 47 (1960).

\bibitem{Gies} 
H. Gies and K. Klingm\"uller,  Phys. Rev. Lett. \textbf{96}, 220401 (2006). 

\bibitem{Bordag} 
M. Bordag and V. Nikolaev, J. Phys. A: Math. Theor. \textbf{41}, 164002 (2008). 

\bibitem{Krause} 
D. E. Krause, R. S. Decca, D. L\'opez, and
E. Fischbach, Phys. Rev. Lett. \textbf{98}, 050403 (2007). 

\bibitem{Smythe} W.R. Smythe, {\it Static and Dynamic Electricity}
(McGraw-Hill, New York, 1968).

\bibitem{Emig1}
T. Emig and R.L. Jaffe, J. Phys. A \textbf{41}, 164001 (2008).

\bibitem{Emig2}
T. Emig, J. Stat. Mech.: Theory Exp., P04007 (2008).

\bibitem{Paulo2008}
P.A. Maia Neto, A. Lambrecht, and S. Reynaud, Phys. Rev. A \textbf{78}, 012115 (2008).

\bibitem{Canaguier} 
A. Canaguier-Durand, P.A. Maia Neto, I. Cavero-Pelaez, A. Lambrecht,
and S. Reynaud, Phys. Rev. Lett. \textbf{102}, 230404 (2009).

\bibitem{ReplyPRARC} 
W. J. Kim, M. Brown-Hayes, D. A. R. Dalvit,
J. H. Brownell, and R. Onofrio, Phys. Rev. A \textbf{79}, 026102 (2009). 

\bibitem{DeccaPFA} 
R. S. Decca, E. Fischbach, G. L. Klimchitskaya,
D. E. Krause, D. L\'opez, and V. M. Mostepanenko, Phys. Rev. D {\bf 79}, 124021 (2009).

\bibitem{Buisseret2007}
F. Buisseret, B. Silvestre-Brac, and V. Mathieu, Class. Quantum Grav. \textbf{24}, 855 (2007).

\bibitem{Giessbl} 
F. J. Giessibl, Rev. Mod. Phys. \textbf{75}, 949 (2003). 

\bibitem{Puppo} 
E. Iacopini, P. Puppo, S. Vettori, and P. Rapagnani,   
in {\sl Frontier Tests of QED and Physics of the Vacuum}, 
edited by E. Zavattini, D. Bakalov and C. Rizzo 
(Heron Press, Sofia, 1998), p. 411.

\bibitem{Derjaguin1934}
B. V. Derjaguin, Kolloid. Z. \textbf{69}, 155 (1934).

\bibitem{Emig}
See section III of T. Emig, A. Hanke, R. Golestanian, and M. Kardar, Phys. Rev. A {\bf 67}, 022114 (2003).

\bibitem{Kim1} 
W. J. Kim, M. Brown-Hayes, D. A. R. Dalvit, J. H. Brownell, and R. Onofrio, 
Phys. Rev. A \textbf{78}, 020101(R) (2008); 
J. Phys. Conf. Ser. \textbf{161}, 012004 (2009).

\bibitem{DeccaComment} 
R. S. Decca, E. Fischbach, G. L. Klimchitskaya,
D. E. Krause, D. L\'opez, U. Mohideen, and V. M. Mostepanenko, 
Phys. Rev. A \textbf{79}, 026101 (2009).

\bibitem{Note} With regard to power-law forces, we would like also to remark 
that, in the Casimir theory-experiment comparison at the basis of the limits to 
non-Newtonian forces, it is also important to  calculate the exact 
Casimir force in the presence of a planar slab of {\sl finite} size, 
especially for the configuration used in \cite{Decca2007}.

\bibitem{Bordag1998}
M. Bordag, B. Geyer, G. L. Klimchitskaya, and V. M. Mostepanenko,
Phys. Rev. D \textbf{58}, 075003 (1998);
\textbf{60}, 055004 (1999); \textbf{62}, 011701(R) (2000).

\bibitem{Masuda} 
M. Masuda and M. Sasaki, 
Phys. Rev. Lett. \textbf{102}, 171101 (2009).

\bibitem{Iannuzzi} 
S. de Man, K. Heeck, and D. Iannuzzi, 
Phys. Rev. A \textbf{79}, 024102 (2009).

\bibitem{Speake} 
C. C. Speake and C. Trenkel, 
Phys. Rev. Lett. \textbf{90}, 160403 (2003).

\bibitem{Kimpatch}
W. J. Kim, A. O. Sushkov, D. A. R. Dalvit, and S. K. Lamoreaux, 
arXiv:0905.3421.

\bibitem{Stipe} 
B. C. Stipe, H. J. Mamin, T. D. Stowe, T. W. Kenny, and D. Rugar, 
Phys. Rev. Lett. \textbf{87}, 096801 (2001).

\bibitem{Kim2} 
W. J. Kim, A. O. Sushkov, D. A. R. Dalvit, and S. K. Lamoreaux, 
Phys. Rev. Lett. \textbf{103}, 060401 (2009).

\bibitem{PPtechnology}
A. Lambrecht, V. V. Nesvizhevsky, R. Onofrio, and S. Reynaud, 
Class. Quant. Grav. \textbf{22}, 5397 (2005).

\bibitem{Bordagpp} 
M. Bordag, G. L. Klimchitskaya, and V. M. Mostepanenko, 
Int. J. Mod. Phys. A \textbf{10}, 2661 (1995).

\end{thebibliography}
\end{document}